\documentclass[floatfix,aps,twocolumn,showpacs,superscriptaddress,prl,10pt]{revtex4-1}
\usepackage{mathrsfs,amsmath,amsfonts,amssymb,graphicx}
\usepackage{epstopdf}
\usepackage{subfigure}
\usepackage{color}
\usepackage{commath}

\begin{document}
\title{Probing quantum turbulence in $^4$He by quantum evaporation measurements}

\author{Ivan Amelio}
\affiliation{Dipartimento di Fisica ``Aldo Pontremoli'', Universit\`a degli Studi di Milano, via Celoria 16, I-20133 Milano, Italy}
\affiliation{Dipartimento di Fisica, Universit\`a di Trento, via Sommarive 14, I-38050 Povo, Italy}
\author{Davide Emilio Galli}
\affiliation{Dipartimento di Fisica ``Aldo Pontremoli'', Universit\`a degli Studi di Milano, via Celoria 16, I-20133 Milano, Italy}
\author{Luciano Reatto}
\email{luciano.reatto@mi.infn.it}
\affiliation{Dipartimento di Fisica ``Aldo Pontremoli'', Universit\`a degli Studi di Milano, via Celoria 16, I-20133 Milano, Italy}

\begin{abstract}
Theory of superfluid $^4$He shows that, due to strong correlations and backflow effects, 
the density profile of a vortex line has the character of a density modulation and 
it is not a simple rarefaction region as found in clouds of cold bosonic atoms. 
We find that the basic features of this density modulation are represented by a wave--packet 
of cylindrical symmetry in which rotons with positive group velocity have a dominant role: 
The vortex density modulation can be viewed as a cloud of virtual
excitations, mainly rotons, sustained by the phase of the vortex wave function.
This suggests that in a vortex reconnection some of these rotons become real so that a vortex 
tangle is predicted to be a source of non-thermal rotons. 
The presence of such vorticity induced rotons can be verified by measurements at low temperature 
of quantum evaporation of $^4$He atoms.
We estimate the rate of evaporation and this turns out to be detectable by current instrumentation.
Additional information on the microscopic processes in the decay of quantum turbulence will 
be obtained if quantum evaporation by high energy phonons should be detected.
\end{abstract}

\maketitle

A unique phenomenon takes place in liquid $^4$He at low temperature: 
quantum evaporation (QE) in which an elementary excitation like a roton or a high energy phonon
impinging on the surface of the superfluid causes the evaporation of a single $^4$He atom \cite{wyatt,V}. 
This phenomenon has given important information on the properties of this strongly interacting Bose system.
In addition, it has been suggested that QE can be used as a probe of other phenomena 
like the detection of solar neutrinos \cite{lanou} and of dark matter \cite{maris}. 
Here we propose that QE can be very useful to uncover aspects of Quantum Turbulence (QT).

QT \cite{A, B} is a paradigm of turbulence that takes place in a pure superfluid, 
i.e. a system in which the normal component 
is essentially zero like in superfluid $^4$He at temperatures well below 1 K.
In QT viscosity cannot play a role 
like in classical turbulence  so other processes must be responsible for the experimentally 
observed \cite{hendry, C} decay of a tangle of quantized vortex lines.

Vortex reconnections in which pairs of vortices intersect and exchange tails 
are relevant processes in a turbulent system to redistribute energy over different length scales
in the most diverse systems, from plasmas of astrophysical or of laboratory interest, to classical or quantum fluids.
Vortex reconnections have a special role in QT because in a superfluid
this is the only mechanism that can change the topology of the vortex tangle generated by an initial forcing.
We have now direct experimental evidence in $^4$He of such reconnection events \cite{D} 
as well of the generation of Kelvin waves \cite{fonda}, 
the elementary excitations of a vortex line \cite{donn}. 
The commonly accepted view of dissipation of energy in QT is based on vortex reconnections 
that excite Kelvin waves and small vortex rings \cite{kursa} and of Kelvin wave cascades that lead to excitations of Kelvin 
waves of larger wave vectors \cite{X, E} until they become efficient phonon 
emitters \cite{F}, so that the vortical energy is dissipated into heat.
There is also theoretical evidence for the direct generation of phonons 
in a vortex reconnection \cite{lead, G}. 
In fact, study of vortex reconnections with the Gross-Pitaevskii equation (GPE) 
has shown that the local merging of the cores of two vortices 
and the following detachment is associated with a shortening of the length of the vortices and with the 
generation of a rarefaction wave that then propagates as phonons. This is a plausible picture but up to now there is no direct 
experimental evidence \cite{H} of the Kelvin wave cascades, of the generation of small vortex rings, or of the rarefaction waves associated to vortex reconnections. 
Therefore fundamental pieces of evidence for the decay of vorticity at very low temperatures 
are still missing.
In the present Letter we present evidence that QE processes \cite{wyatt}
should be induced by a vortex tangle due to vortex reconnections thus giving microscopic 
insight into the decay of QT.
In fact, we find that 
the vortex core structure given by state of the art quantum many-body simulations \cite{prb14}
can be recovered as a cylindrically symmetric wave--packet (WP) of bulk roton states, suggesting the picture 
of the vortex as a cloud of virtual excitations, mainly rotons, induced by the flow field.
This leads us to the conjecture that part of the energy from reconnection events
is in the form of non-thermal rotons.
We estimate the rate of roton emission from a tangle, and show that these rotons should be detectable \cite{lt_reatto} via processes of QE of ${}^4$He atoms \cite{wyatt}, if the liquid has a free surface.
QE should also provide information on the Kelvin cascade in the high-energy phonon region.

\begin{figure*}[t]
\centering
\includegraphics[width = 0.95\textwidth]{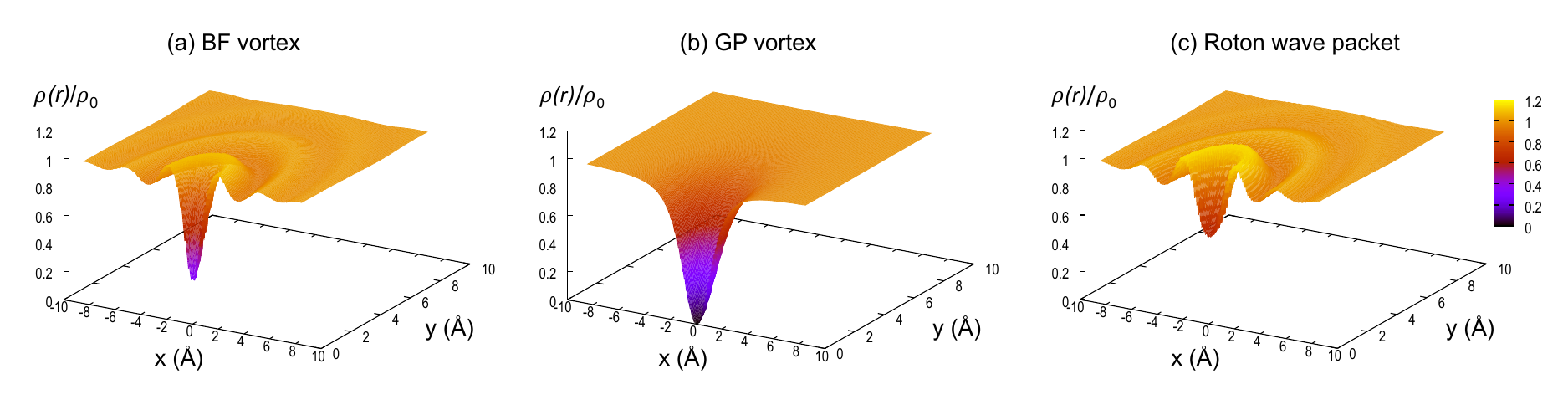}
\caption{ Rescaled density $\frac{\rho(x,y)}{\rho_0}$ of the (cylindrically symmetric) vortex, 
where the vortex axis is along $z$,  as computed from (a) BF--SPIGS \cite{prb14},
(b) GPE with coherence length $\xi=0.87$ \AA$^{-1}$ \cite{zucc}. 
In (c) the rescaled density for the wave packet (1) with parameters 
$A_1=-0.8$, $\sigma_1=0.25$ \AA$^{-1}$, $q_2=1.95$ \AA$^{-1}$, $\sigma_2=0.35$ \AA$^{-1}$ is shown.}
\label{fig:profile} 
\end{figure*}

The theoretical efforts to study QT are based on phenomenological 
Biot--Savart models \cite{saff, A} or on the mean field approximation as embodied in the GPE. 
While GPE gives a very accurate description of cold bosonic atoms and its predictions on vorticity 
in clouds of such atoms have been beautifully verified experimentally \cite{L}, 
it is known that GPE gives a very poor representation of superfluid ${}^4$He.
For instance the excitation spectrum $\epsilon(q)$ given by GPE is a crossover from the
phonon region at small wave vector $q$ 
to a free particle $q^2$ behavior at large $q$ \cite{W}, so it misses completely 
the maxon-roton feature so characteristic of superfluid ${}^4$He \cite{Z}.
The GPE static density response function $\chi_\rho (q)$
is a Lorentzian function of $q$ centered at $q=0$, 
a behavior completely different from the experimentally determined $\chi_\rho (q)$ 
that is characterized by a sharp peak at $q \simeq 2.$ \AA$^{-1}$ \cite{Z}.
Since long time it is known from many--body computations \cite{ches, sadd} that
the short range structure of the vortex core in ${}^4$He 
is much more complex of the simple rarefaction region \cite{W} given by GPE, 
in which the local density $\rho(r)$ 
vanishes at the vortex axis $r=0$ and smoothly approaches the bulk density at large distances.
As a result, we expect the GPE to provide plausible conclusions for the large scale dynamics of the vortex tangle, while phenomena like vortex reconnections, requiring the full treatment of strong correlations
at atomic length scale, need further scrutiny.

The recent many--body computation \cite{prb14} of a vortex line in liquid ${}^4$He at $T=0$ K 
is based on the fixed phase approximation:
by writing the vortex wave function $\psi_v (R)$, $R=(\vec{r}_1, ..., \vec{r}_N)$, 
in term of its modulus and phase,
$\psi_v (R) = \abs{\psi_v (R)} \exp \left[ i \Phi(R) \right]$, one makes an ansatz 
for the functional form of $\Phi(R)$,
obtaining a Schr\"odinger like equation for $\abs{\psi_v (R)}$ \cite{orti}.
This equation was solved \cite{prb14} by Shadow Path Integral Ground State 
(SPIGS) \cite{spigs1,spigs2} Monte Carlo simulation, an unbiased ``exact'' method \cite{lesse}.
The resulting local density $\rho(r)$ is not a monotonic function of the distance $r$ 
from the vortex axis and it approaches the 
bulk density, $\rho_0$, in an oscillating way (see Fig.~\ref{fig:profile}).
The best vortex energy is obtained when the phase $\Phi(R)$ contains backflow terms
(i.e. terms depending on positions of pairs of particles) and one finds
three related features \cite{fett}: the density $\rho(r=0)$ on the axis is non-zero, 
the velocity field $\vec{v}(\vec{r})$ 
at short distance deviates from the $r^{-1}$ behavior given by GPE with $\vec{v}(\vec{r})$ being finite 
even at $\vec{r}=0$ and $\nabla \times \vec{v}(\vec{r})$ is non zero in a finite region around 
the vortex axis \cite{sadd}.

\begin{figure}[b]
\centering
\includegraphics[width = 0.45\textwidth]{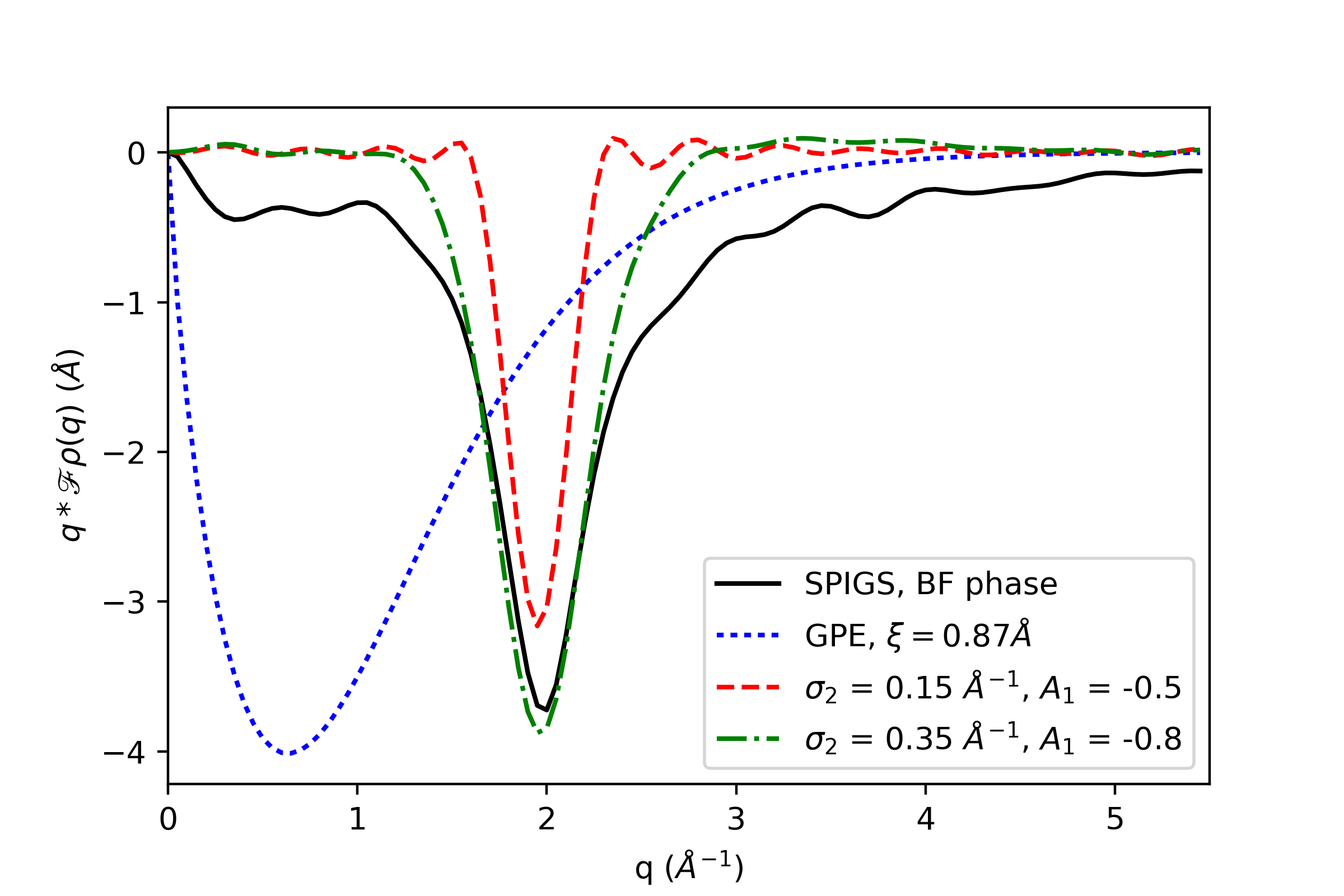}
\caption{Fourier transform times $q$, $q \int dr \ r J_0(qr) \Delta\rho(r)$ ($J_0$ being a Bessel function),
of the (cylindrically symmetric) density variation, $\Delta\rho(r)$, of a vortex as computed from GPE with 
coherence length $\xi=0.87$ \AA$^{-1}$ (blue dotted), of a BF--SPIGS vortex (solid black) and of 
two wave packets with parameters
$A_1=-0.5$, $\sigma_1=0.25$ \AA$^{-1}$, $q_2=1.95$ \AA$^{-1}$, $\sigma_2=0.15$ \AA$^{-1}$ (red dashed)
and $A_1=-0.8$, $\sigma_1=0.25$ \AA$^{-1}$, $q_2=1.95$ \AA$^{-1}$, $\sigma_2=0.35$ \AA$^{-1}$ (green dot-dashed);
this last one has the density profile shown in Fig.~\ref{fig:profile}(c).}
\label{fig:tdf} 
\end{figure}

It is instructive to look not only at $\rho(r)$ but also at the Fourier transform,
$\mathcal{F}\rho(q)$,
of the adimensional density variation $\Delta\rho(r)=\rho(r)/\rho_0-1$
(here and in the following we use the convention that momenta $\vec{q}$ lie in the $xy$-plane).
$\mathcal{F}\rho(q)$ multiplied by $q$ at the equilibrium density of 
${}^4$He is shown in Fig.~\ref{fig:tdf} for the SPIGS
computation with the backflow phase \cite{ftnota}, as well as the result for the GPE.
The GPE $q\mathcal{F}\rho(q)$ has a rather wide minimum at $q$ in the phonon region 
whereas the fixed phase $q\mathcal{F}\rho(q)$
is rather small in the phonon region and is dominated by a sharp minimum at a larger $q_{min}$.
At the equilibrium density $q_{min} \simeq 2.0$ \AA$^{-1}$ is very close to the position of the peak of the static density response function
$\chi_\rho(q)$ and somewhat larger of $q=1.91$ \AA$^{-1}$, the wave vector of the roton minimum.
This behavior has been verified at all densities in the fluid phase \cite{prb14}.
Thus the spectrum of the density profile is dominated by wave vectors corresponding to $R^+$ rotons, rotons with positive group velocity.

In classical hydrodynamics of an incompressible fluid two antiparallel vortex lines 
form a stable object. 
In the quantum case the behavior is quite different as shown by GPE: 
due to the finite quantum compressibility a pair of antiparallel vortices approach each other until 
the phases of the two merging vortices annihilate, leaving a rarefaction region \cite{G}, 
which expands and propagates as phonon excitations.
A more general model of the reconnection dynamics, consisting in two intersecting vortex rings \cite{lead} 
as well for generic shape \cite{U}, indicates a shortening of the vortex line length with 
formation of a rarefaction region.
We can understand this GPE result as a way of avoiding sharp kinks of the two vortices after 
reconnection because this would correspond to a very highly excited state of Kelvin waves.
The process of avoiding high curvature cusps in the vortex system
is expected to be generic so it should happen also in a strongly interacting system like $^4$He.
In this case, however, what is left after the local phase annihilation of the two reconnecting vortices is not a rarefaction region
but a density modulation dominated by wavevectors of order of 2 \AA$^{-1}$.
This modulation is no more sustained by the centrifugal force associated with the phase $\Phi(R)$ 
and can be efficiently described in terms of bulk excitations, that will propagate carrying away some energy.

In order to get insight into the nature of the excitations generated in a reconnection, 
we pose the following question: which wave--packets,
$\psi(R)= \int d\vec{q} \pi(\vec{q}) \psi_{\vec{q}}(R)$, built up from the {\em bulk}
single--excitation states
$\psi_{\vec{q}} $, yield a cylindrical density modulation with similar features to those given by
$\abs{\psi_v (R)}$?
A standard WP is centered around a given wave vector 
$\vec{q}$ and position $\vec{r}$. 
In order to have a packet with cylindrical symmetry with respect to the $z$ axis $\vec{q}$ has 
to be normal to the vortex axis and it has to be averaged over the directions in the $q_x-q_y$ plane.
In addition, one has to average also with respect to the directions of $\vec{r}$ in the $x-y$ plane 
if $\vec{r}$ does not lie on the vortex axis.
See Supplemental Material (SM) for such averages.
At the end one can write the packet as $\psi(R)= \int d\vec{q} \pi(q) \psi_{\vec{q}}(R)$.
Thus we restrict to packets of cylindrical symmetry $\pi (\vec{q})=\pi(q)$, $q=\sqrt{q^2_x + q^2_y}$, and as 
wave function of the bulk excited states we adopt for $\psi_{\vec{q}}$ 
the shadow variational wave function \cite{N, macf, cecc, moroni} and explore different packets, 
as discussed in the SM.
Density can be computed by means of a Monte Carlo sampling and search for parameters of the packets 
giving density profiles close to that of the vortex.
{In Fig.\ref{fig:profile} one profile is shown}
corresponding to a double gaussian $\pi(q)$:
\begin{equation}
\pi_{ \{A_1,\sigma_1,q_2, \sigma_2 \} } (q) = 
A_1 e^{-\frac{q^2}{2 \sigma_1^2}} + e^{-\frac{(q-q_2)^2}{2\sigma_2^2}}
\label{eq:bigauss}
\end{equation}
one centered at $q=0$ and one in the roton region.
One can notice that the shape of the vortex density profile is well reproduced by our roton WPs, 
the deviations are presumably due to multiple excitations not included in our model (see SM). 
The amplitude of the density oscillations of the WPs depends on the length $L_z$ of the simulation box. 
By changing $L_z$ the shape of $\mathcal{F}\rho(q)$ remain essentially unchanged but its amplitude scale
roughly as $1/L_z$ because the effect of our single excitation WP is spread over a region proportional 
to $L_z$ (see SM). 
In order that the amplitude of the density oscillation of the WP matches that of the vortex 
as in Fig.~\ref{fig:tdf} one excitation per about 25-30 \AA~ is needed.
There is a nice consistency check of this because the contribution of $\abs{\psi_v (R)}$
to the vortex energy is 0.4 K/\AA~ (see SM) 
so that a length of order of 25 \AA~ corresponds to the energy of a roton.
Computing the energy density as $ q|\pi(q)|^2\epsilon (q)$, we find that R$^+$ rotons with wavevector 
$q \in [1.93, 2.15] $ account for the $30$-$50 \% $ of total energy.
The R$^+$ contribution might even be higher due to multiple 
excitation contributions (see SM).
Using WP with only phonons (maxons), we obtain density profiles with very weak modulations (different oscillation wavelengths).
The shown results are robust to changing Gaussian into Lorentzian or to adding a third peak so we conjecture that the dissipation waves emitted in a reconnection event have a low-energy phonon component plus an energetically relevant roton contribution.

\begin{figure}[tp]
\centering
\includegraphics[width = 0.95\columnwidth]{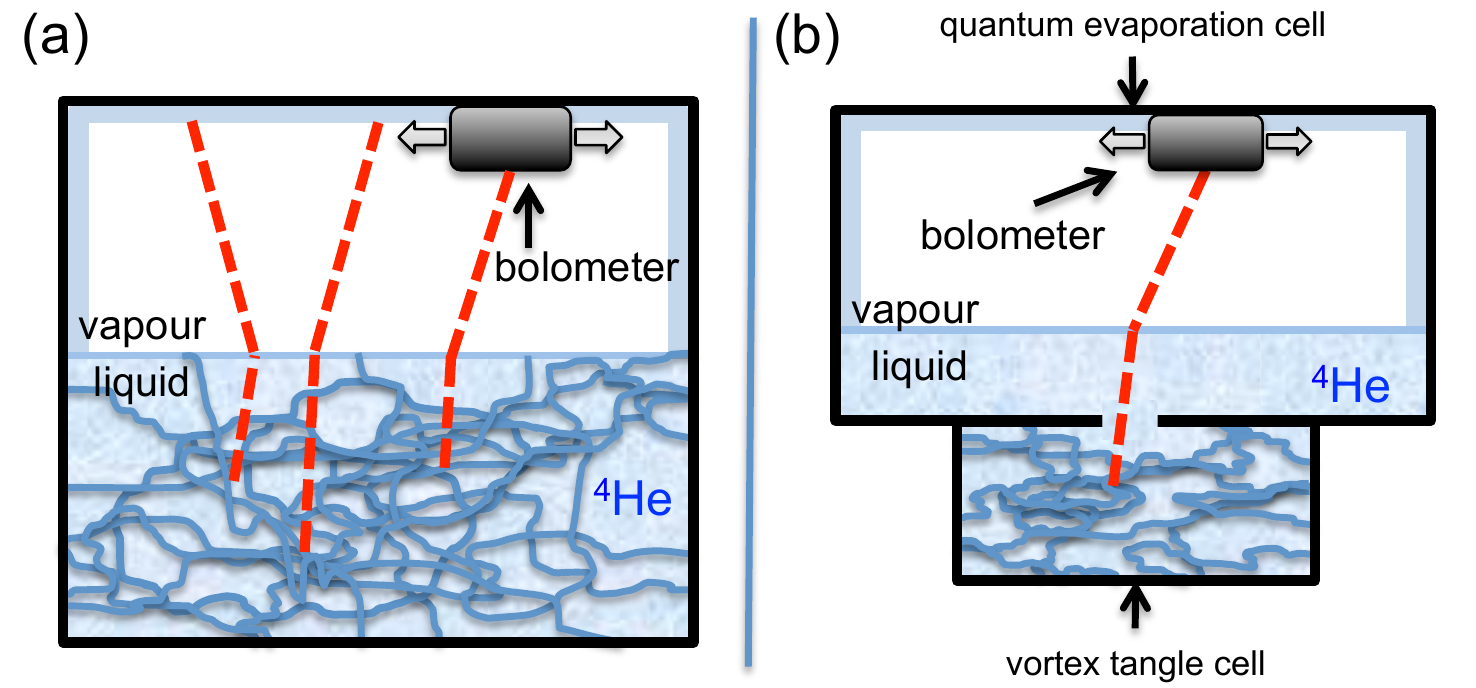}
\caption{Quantum evaporation setup to detect excitations generated in vortex reconnections.
Red dashed lines represent excitations (rotons) transforming into evaporated $^4$He. 
Scheme (a) is suited for measuring the dependence of the signal on the amount of turbulence $L$, 
while (b) is for an energy-resolved detection of particles.}
\label{fig:apparato} 
\end{figure}

The presence of energetic rotons in the superfluid even at very low T due to vortex reconnections 
can be experimentally detected because such rotons will be able to cause QE of 4He atoms \cite{wyatt}: 
QE can take place under the conditions that the excitation propagates ballistically in the bulk, that 
its energy is larger than $E_b=7.15$ K, the binding energy of $^4$He in the liquid,
with conservation of momentum parallel to the interface and
of energy, $\epsilon_q = E_b + \hbar^2 k^2 / 2m $, 
where $\hbar\vec{k}$ is the momentum of the $^4$He evaporated atom.
QE has been detected for phonons of energy above about 10 K, 
for R$^-$ and for R$^+$ rotons, with R$^+$ rotons having the largest efficiency for QE, of order of 0.3.
At $T \ll 1$ K the number of thermal rotons and high energy phonons is negligible so no evaporation 
of $^4$He atoms should take place.
On the basis of our conjecture, a vortex tangle is a source of non-thermal rotons, so that one should observe 
processes of QE even at low $T$, as long as the tangle is present in a superfluid that has a free surface 
(see Fig.\ref{fig:apparato}a). 
In order to prove that QE from a vortex tangle is due to rotons, one needs to focalize the excitations 
and to perform time of flight measurements, so that one can use the conservation laws
to verify the dynamics of the process. 
A possible experimental setup is shown in Fig.\ref{fig:apparato}b
with one chamber where vorticity is generated and an evaporation upper chamber only partially filled with $^4$He 
and with the detecting bolometers. The two chambers are connected by a periodically opened duct or the generation 
of the vortex tangle is periodic in time.

A crucial aspect is if the rate of evaporated atoms is large enough to be detected. 
We can evaluate this rate starting from the frequency of reconnections per unit volume
in a random tangle \cite{frec}
\begin{equation}
f_{rec} = \frac{\kappa}{6 \pi} L^{5/2} \ln(L^{-1/2}/a_0) \label{eq:4}
\end{equation}
where $\kappa = h/m \simeq 10^{-3}$ cm$^2$/s is the quantum of circulation, $a_0$ is of order 1 \AA~
and $L$ is the total length of vortex lines per unit volume.
Typical experimental values of $L$ are in the range $10^2$-$10^6$ cm$^{-2}$. 
For such values of $L$, the volume of the cores of vortices is negligible, 
so rotons should propagate ballistically through the tangle.
The logarithmic term in eq.(\ref{eq:4}) depends very weakly on $L$ in this range and
$\ln(L^{-1/2}/a_0) / 6 \pi$ is close to unity.
For example for $L=10^2$, $10^4$ and $10^6$ cm$^{-2}$ we get, respectively, 
$f_{rec} \simeq 10^2$, $10^7$ and $10^{12}$ cm$^{-3}$s$^{-1}$.
Next we need an estimate of how many rotons are emitted in a reconnection.
In GPE the energy $\Delta E$ transferred from vortex flow energy to the rarefaction wave depends 
on the geometry of the reconnecting vortices and $\Delta E = 10$ K 
is the typical value \cite{bar_new}.
As an order of magnitude estimate we can assume this GPE value for $\Delta E$ also for $^4$He 
because the GPE vortex energy with coherence length 0.87 \AA~ as in Ref.~\cite{zucc} is in good 
agreement with the many-body computation also at short distance (see SM).
Estimating which percentage of this energy is dissipated in rotons rather than in phonons
is a main goal of the experiment.
Taking 10\% as a lower bound for roton emission, using the known probability 0.3 for 
quantum evaporation by R$^+$ rotons \cite{O, P} and the about 5\% probability that the roton 
impinges on the surface within $25^{\circ}$ from the vertical so that it can give QE \cite{maris},
we get that $f_{ev}$, 
the rate of evaporated atoms per unit time and unit volume of the tangle, 
is $f_{ev} \simeq 10^{-1}$, $10^{4}$ and $10^9$ cm$^{-3}$s$^{-1}$
for $L=10^2$, $10^4$ and $10^6$ cm$^{-2}$ respectively.
A bolometer of sensitivity $10^{-11}$ erg \cite{Q} is able to detect the energy of about 
$10^4$ rotons so the estimated number of evaporated atoms should be detectable with current 
instrumentation \cite{footnote2}, at least for $L > 10^4$ cm$^{-2}$.

The scaling of measured evaporated $^4$He versus $L$, which can be independently measured \cite{C}, 
allows a consistency check for eq. (\ref{eq:4}) with the fact that the measured evaporated atoms 
originated from reconnections.
An additional reason of interest for performing QE experiments in presence of a vortex tangle is to assess the presence of high energy phonons in quantum turbulence. By high energy phonons we mean phonons with 
$q>k_c \simeq 0.55$ \AA$^{-1}$ ($\epsilon_q \gtrsim 10$ K), where at $k_c$ 
the dispersion changes from anomalous to normal in the liquid at s.v.p. \cite{R}, 
so that such phonons do not decay spontaneously, but propagate ballistically  and can produce QE. 
Present theories of dissipation in QT predict that phonons should be emitted with $q$ well below $k_c$.
For $^4$He there is really no quantitative microscopic theory 
of Kelvin waves and of their interaction with phonons at large $q$, so that detection of processes of QE 
by non-thermal high-$q$ phonons from a vortex tangle would add important information on such aspects.

In summary, advanced quantum simulations of a vortex line in ${}^4$He and of roton in the bulk show 
that the vortex core structure can well be represented by a cylindrical WP of rotons 
so we can view the vortex core mainly as a cloud of virtual rotons sustained by the phase of the vortex
wave function.
In a natural way this leads to the conjecture that 
in a vortex reconnection some of the virtual rotons become real and propagating.
We estimate the number of such non-thermal rotons and of the expected rate of 
quantum evaporation processes that should be large enough to be detected with current instrumentation.
Additional information on quantum vorticity will be obtained if the quantum evaporation measurement should detect also 
the presence of high energy phonons. With such experiment one could give light on some of the microscopic processes 
that are important in the evolution of a vortex tangle in a pure superfluid.
On the theory side it will be interesting to study vortex reconnections with a time--dependent 
non--local density functional like that in Ref.\cite{anci} that gives a good description of the vortex structure in $^4$He.

%

\pagebreak
\onecolumngrid
\vspace{\columnsep}
\newpage
\begin{center}
\textbf{\Large Supplemental Material: \\
Probing quantum turbulence in $^4$He by quantum evaporation measurements}
\\
\vspace{0.5cm}
Ivan Amelio$^{1,2}$, Davide Emilio Galli$^{1}$, Luciano Reatto$^{1}$
\\
$^{1}$ Dipartimento di Fisica ``Aldo Pontremoli'', Universit\`a degli Studi di Milano, via Celoria 16, I-20133 Milano, Italy
\\
$^{2}$ Dipartimento di Fisica, Universit\`a di Trento, via Sommarive 14, I-38050 Povo, Italy
\\
\end{center}
\vspace{2cm}
\twocolumngrid

\setcounter{equation}{0}
\setcounter{figure}{0}
\setcounter{table}{0}
\renewcommand{\theequation}{S\arabic{equation}}
\renewcommand{\thefigure}{S\arabic{figure}}
\renewcommand{\thetable}{S\arabic{table}}
\renewcommand{\bibnumfmt}[1]{[S#1]}
\renewcommand{\citenumfont}[1]{S#1}
\addtolength{\textfloatsep}{1mm}

\section{A) Vortex Excitation Energy}

\begin{figure}[b]
\centering
\includegraphics[width = 0.45\textwidth]{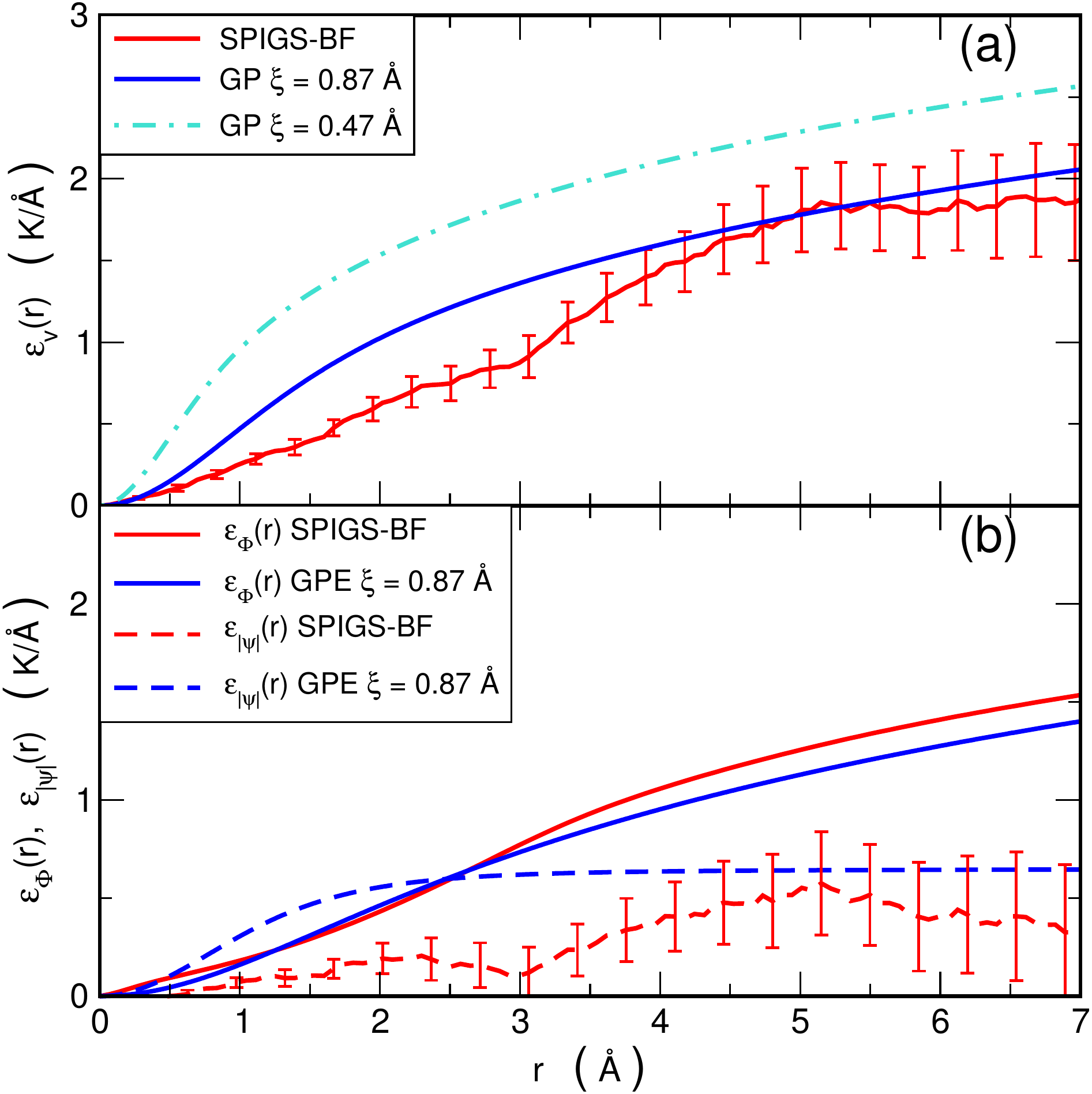}
\caption{(a) Comparison of integrated energy up to distance $r$ of a single vortex line per unit 
 length, $\varepsilon_v(r)$, at equilibrium density obtained with different methods.
 (b) Comparison of the expectation value of the kinetic energy due to the phase, $\epsilon_\Phi(r)$,
 and of $\epsilon_{\abs{\psi}}(r)$ at equilibrium density obtained with different methods.}
\label{fig:new_SM} 
\end{figure}

In Fig.~\ref{fig:new_SM}(a) we show the excitation energy per unit length of a straight vortex line 
in $^4$He at the equilibrium density integrated up to the radial distance $r$ from the vortex 
axis for the SPIGS--backflow wave function \cite{prb14S} and for the Gross-Pitaevskii equation (GPE) for two values of the 
coherence length. The vortex excitation energy $\epsilon_v(r)$ can be decomposed into several
contributions and it is instructive to see how they depend on $r$.
One contribution is $\epsilon_\Phi(r)$, the expectation value of the kinetic energy 
due to the phase $\Phi$ of the wave function.
Another contribution represents the extra kinetic energy due to the bending of the real part 
of the wave function close to the core. Finally there is a contribution due to the change
of the expectation value of the interparticle interatomic potential due to local rearrangement 
of the atoms close to the core as a consequence of the inhomogeneity induced by the phase. 
We call $\epsilon_{\abs{\psi}}(r)$ the sum of these two last terms.
A similar decomposition can be performed for the GPE vortex energy.
In Fig.~\ref{fig:new_SM}(b) $\epsilon_\Phi(r)$ and $\epsilon_{\abs{\psi}}(r)$ are plotted 
as function of $r$ for the SPIGS--backflow wave function and for GPE. 
It can be noticed that $\epsilon_{\abs{\psi}}(r)$ rapidly saturates to a plateau value of around $0.4$ K/\AA,
so that beyond a distance of order of 4 \AA~ the increment in $r$ of the vortex energy 
is only due to the phase of the wave function.
In the case of SPIGS-backflow this plateau is around 0.4 K/\AA~ while the GPE value is 0.65 K/\AA. 
It is interesting to notice that the SPIGS computation \cite{prb14S} with the Onsager--Feynman phase, 
i.e. the same phase of GPE, gives a plateau value of around 0.8 K/\AA.
Therefore this vortex core energy is quite sensitive to the chosen phase and it is strongly reduced by backflow correlations.

\section{B) Excited state wave function}

\begin{figure}[b]
\centering
\includegraphics[width = 0.48\textwidth]{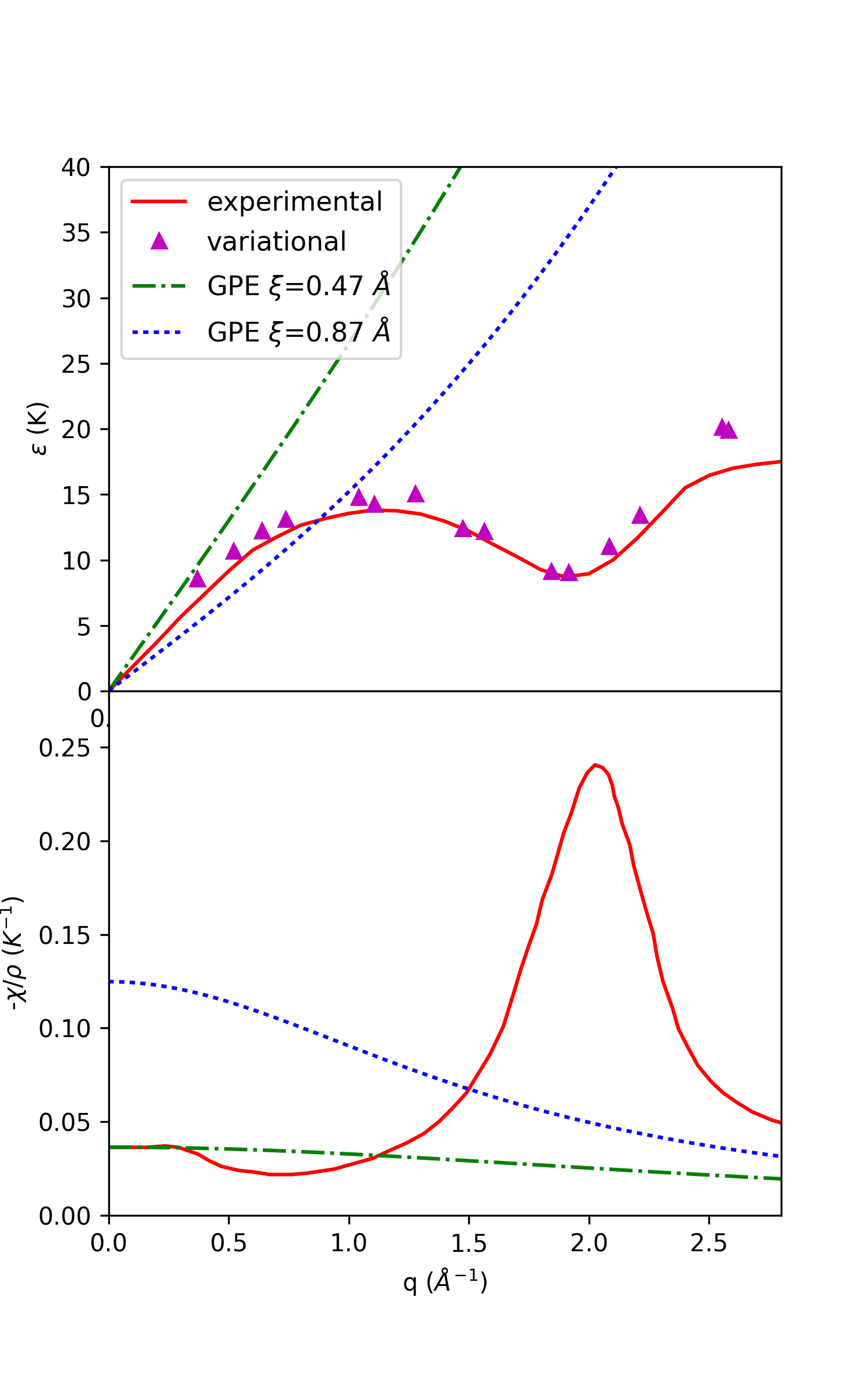}
\caption{{(top) Schematics of the dispersion law for Helium, as found experimentally (solid red), 
by Gross-Pitaevskii with two different healing lengths, $\xi = 0.47$ \AA~, chosen to fit 
short-wavelength properties, (green dot-dashed) and $\xi = 0.87$ \AA~ (blue dotted), obtained 
fitting the core parameter \cite{prb14S}, which determines the long scale behaviour of the vortex energy.
Finally, the triangle data is the energy computed in \cite{ceccS} by sampling the variational 
wavefunction (\ref{eq:psi_shadow}).  (bottom) Static density response function.}}
\label{fig:spettro} 
\end{figure}

\begin{figure*}[t]
\centering
\includegraphics[width = 0.48\textwidth]{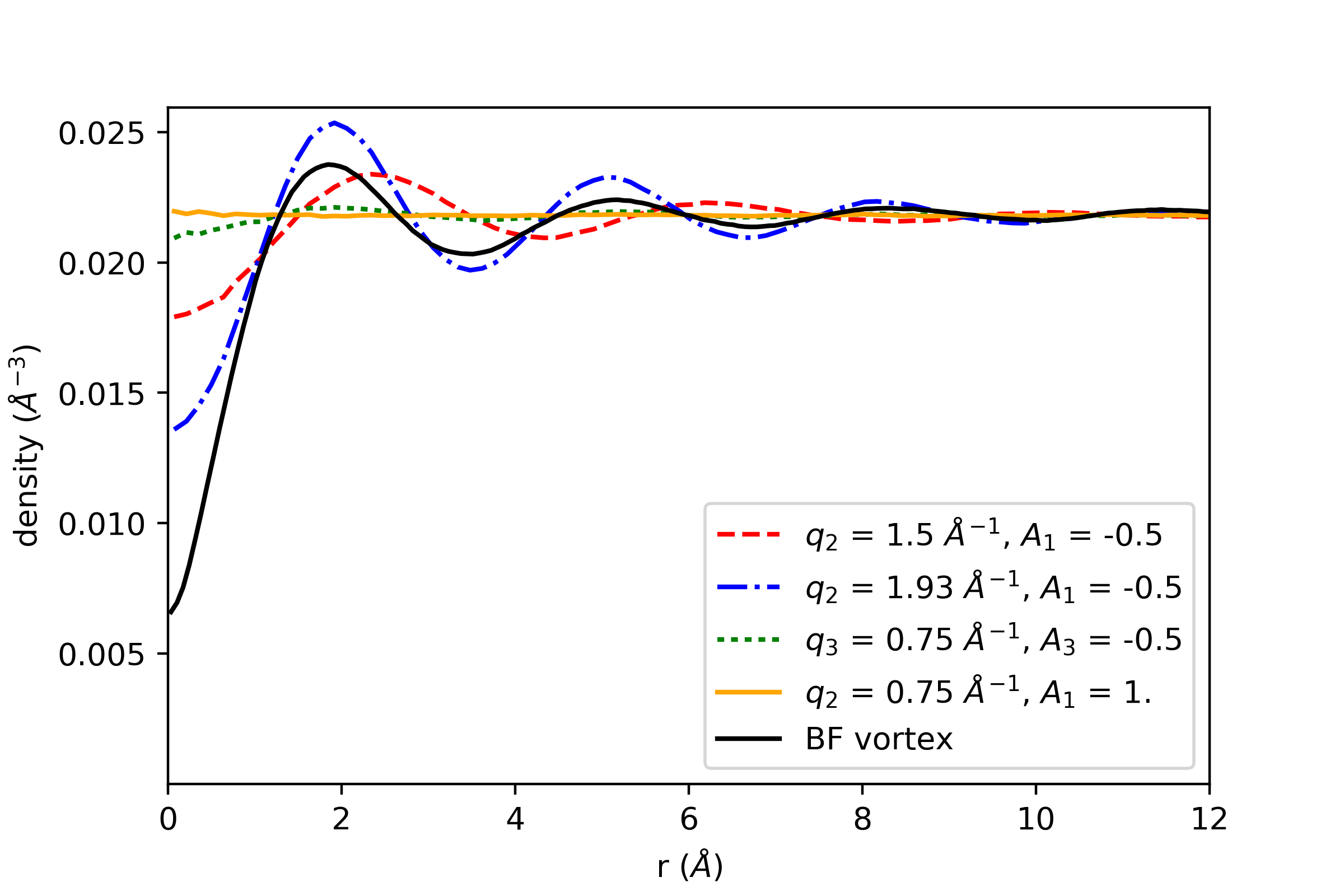}
\includegraphics[width = 0.48\textwidth]{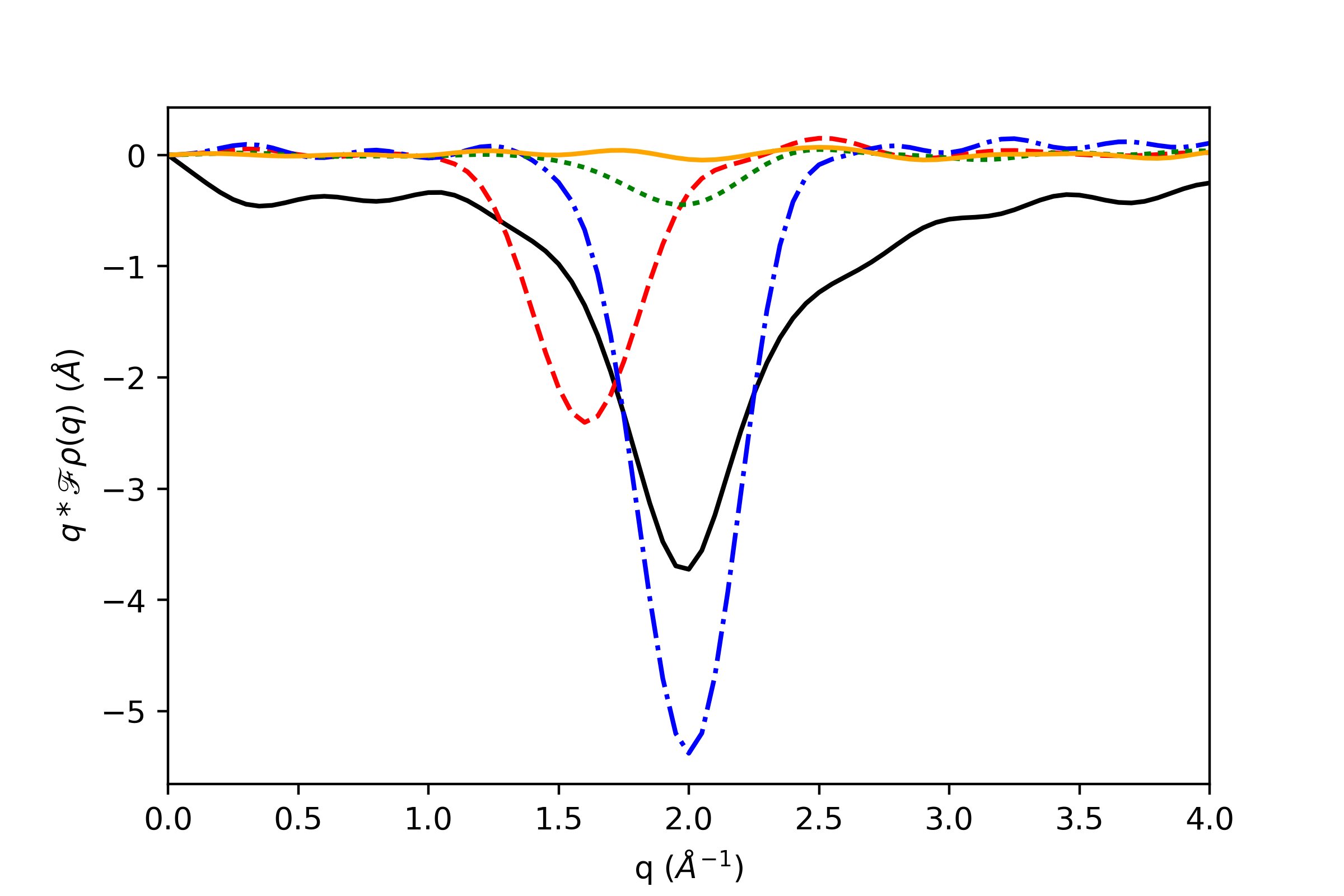}
\caption{(left panel) Density of the (cylindrically symmetric) vortex as a function of the radial 
distance from its axis and (right) Fourier transform times 
$q$, $q \int dr \ r J_0(qr) \Delta\rho(r)$ ($J_0$ being a Bessel function),
of the (cylindrically symmetric) density variation, $\Delta\rho(r) = \rho(r)/\rho_0 -1$, as computed from BF-SPIGS (solid black) and some bigaussian packets, see eq. (\ref{bigauss}). In the green dotted plot, a third gaussian has been added to the packet corresponding to the blue dash--dotted line. The parameters are reported in figure, the $\sigma_i$ being equal to $0.25$ \AA$^{-1}$ in all the cases.}
\label{fig:profileS} 
\end{figure*}

\begin{figure*}[t]
\centering
\includegraphics[width = 0.48\textwidth]{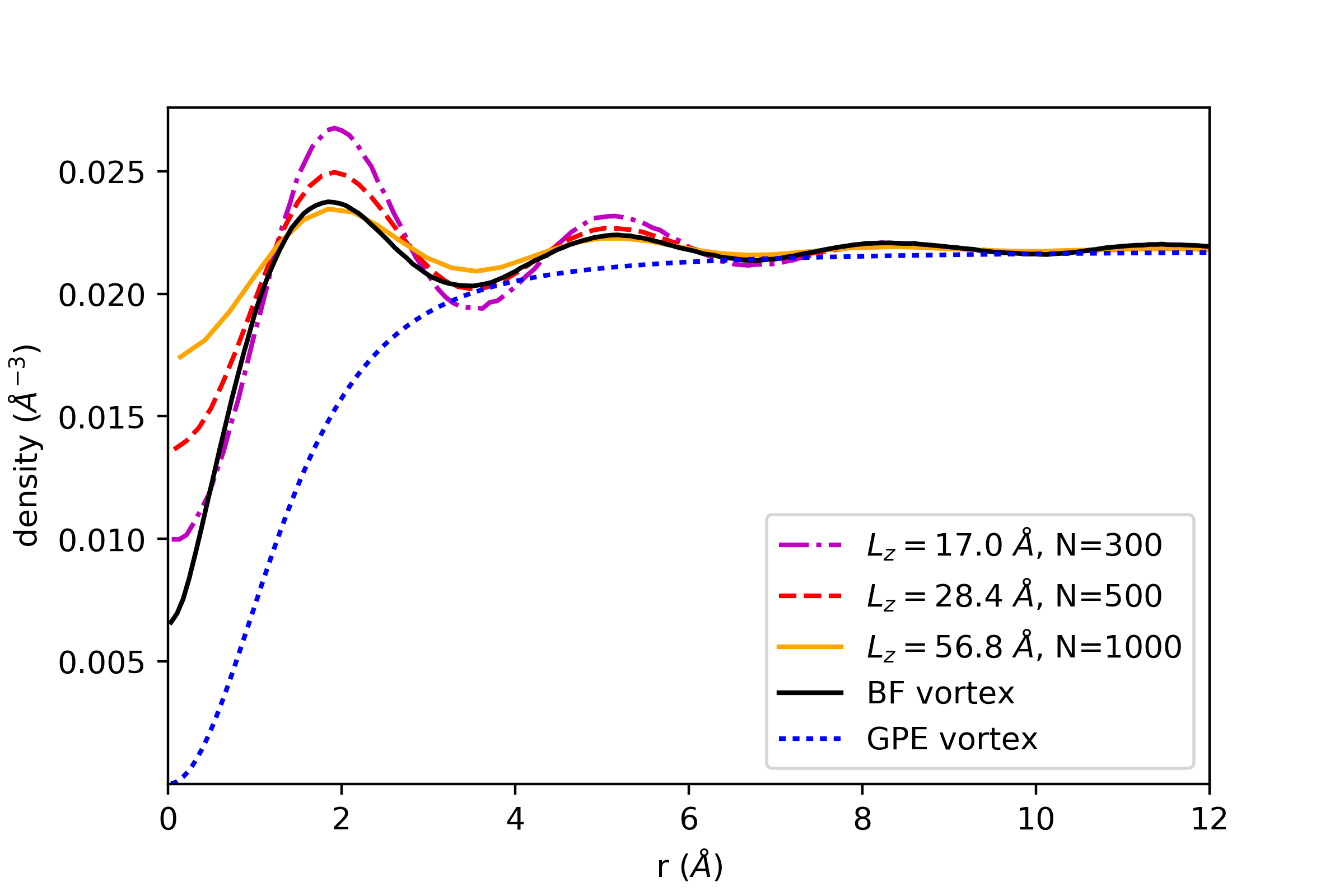}
\includegraphics[width = 0.48\textwidth]{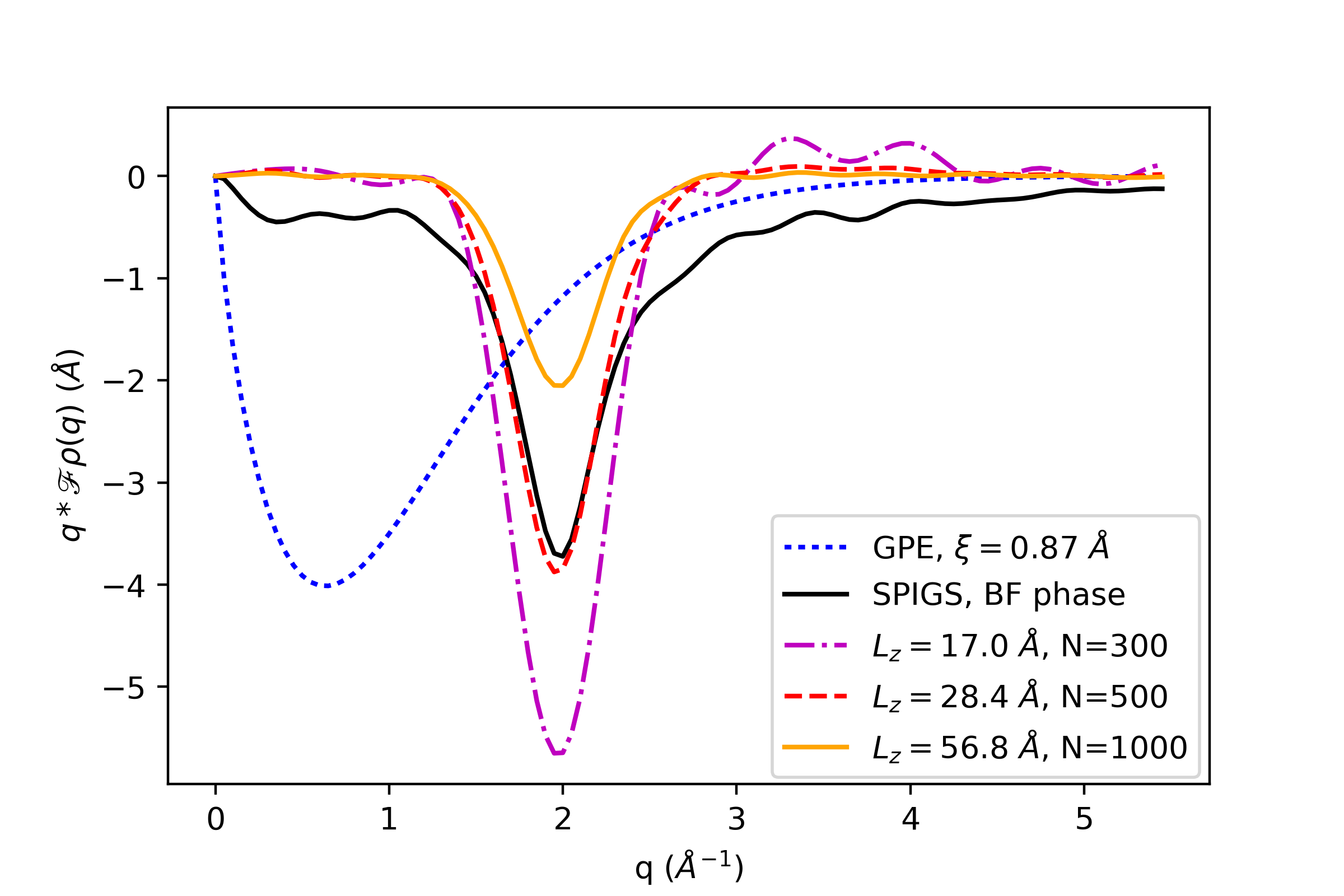}
\caption{Scaling of the (left panel) density and of its (right)  Fourier transform  
 $q \int dr \ r J_0(qr) \Delta\rho(r)$ for different values of $L_z$, i.e. of how much a single many-body excitation is diluted. Notice that in this computations the mean density is constant, since $L_z \propto N$. The parameters of the packet are the same as the packet shown in the main text, that is $q_2=1.95$ \AA$^{-1}$,
$\sigma_1=0.25$ \AA$^{-1}$, $A_1=-0.8$, $\sigma_2=0.35$ \AA$^{-1}$.}
\label{fig:N_Scaling} 
\end{figure*}

In order to build the {single-excitation} wave packet
\begin{equation}
\Psi (R) = \int d^2 q \ \pi (\vec{q}) \psi_{\vec{q}} (R)
\label{eq:packet}
\end{equation}
(here and in the following we use the convention that momenta $\vec{q}$ lie in the $xy$-plane, normal to the vortex axis) we need an explicit form for the excited states
 wave function $\psi_{\vec{q}} (R)$. 
We  adopt the shadow variational scheme \cite{mcfar, ceccS} which approximates 
the many-body wavefunction of an elementary excitation of momentum $\vec{q}$ as:
\begin{equation}
{\psi}_{\vec{q}}(R) = \frac{1}{\sqrt{N_q}} \int dS \ F(R,S) \sum_j e^{i\vec{q}\cdot \tilde{\vec{s}}_j}
\label{eq:psi_shadow}
\end{equation}
where ${R=\{ \vec{r}_j \} }_{j=1}^N$ are the particles coordinates, the "shadow" integration variables $S=\{ \vec{s}_j \}_{j=1}^N$  are a way to expand the variational space, while we introduce 
\begin{equation}
\tilde{\vec{s}}_j = \vec{s}_j + \sum_{i \neq j} \beta(q) \lambda (|\vec{s}_i - \vec{s}_j|)(\vec{s}_i - \vec{s}_j) \label{eq:backflow}
\end{equation}
as in \cite{ceccS} to take into account explicit backflow effects. 
$F(R,S)$ is in the  form $F(R,S) = \prod_{i<j} f_{rr}(|r_i-r_j|) \prod_{k<l} f_{ss}(|s_k-s_l|) \prod_{m} f_{rs}(|s_m-r_m|)$,
with $f_{rr}$, $f_{rs}$ and $f_{ss}$ non negative Jastrow functions, that are determined by the variational principle of minimizing the ground state energy $E(0)$. $N_q$ is the normalization constant.

The energy of an excitation $\epsilon (q)$ is defined by $\epsilon (q)=E(q)-E(0)$, $E(q)$ being the energy of (\ref{eq:psi_shadow}). This expectation value, as well as other observables, can be computed as the average of a proper observable in the 
space of $3 \times N$ position variables, corresponding to a (classical) Metropolis dynamics on the probability distribution $F^*(R,S_1)F(R,S_2)$. The simulation is carried out in a box with periodic boundary conditions: to this end we constrain the one--particle shadow wavefunction to be zero outside a cylinder of radius $s_{max}/2$ contained in the simulation box, and to impose that in a continuous way we replace 
$\eta_0(
{|\vec{s}}_j|) \equiv \int  \frac{d^2 q}{\sqrt{N_q}} \ \pi (\vec{q}) e^{i\vec{q}\cdot 
{\vec{s}}_j}$ with $\eta(s_j) \equiv \eta_0(
{{s}}_j ) + \eta_0(
{{s}}_j - s_{max}) - 2 \eta_0(
{ s_{max}}/2)$ for $s<s_{max}/2$ and $0$ otherwise.
If $\pi(q)$ is not too peaked, both the wavefunction and its density profile have been checked to be insensitive to changes in $s_{max}$, due to the exponential decay of $\eta(s)$ (see multi--modal form of $\pi$ below).
Where not otherwise indicated, we report results for $N=500$ particles.
  
Minimizing with respect to the variational parameters contained in (\ref{eq:backflow}), $\epsilon (q)$ is in good quantitative agreement with
the excitation spectrum of superfluid $^4$He, achieving even in the roton region an error not greater 
than 5\% . 
{This is illustrated in Fig. \ref{fig:spettro}, where it is also recalled that Gross--Pitaevskii mean--field completely misses the rotonic feature; we report also the static density response function, which we commented on in the main text.}
In particular, the optimal value for the backflow strength, $\beta(q)$,
depends explicitly on $q$, ranging from 0 for phonons to 0.3 for rotons. 
We have not used the fully optimized \cite{moroniS} Jastrow functions in (\ref{eq:psi_shadow}) but we have used 
the simplified form of Ref.~\onlinecite{mcfar} because the (marginal) improvement in the energy is not important 
for the present purpose of computing wave packets.

\section{C) Cylindrical wave packets}

We restrict to wave packets of cylindrical symmetry $\pi (\vec{q}) = \pi (q)$ with respect to the $z$--axis. Initially, the physical picture led us to packets built from bulk states  localized at a radius $r_0$ around the vortex and distributed, in momentum space, according to a gaussian around $\vec{q}_R$, $\vec{q}_R$ being averaged on the plane:
\begin{equation}
\Psi_{\Delta, q_R, r_0} (R) = \int_0^{2 \pi} d\theta_q d\theta_r \hat{T}_{\vec{r}_0} \int d^2 k e^{-\Delta ({\vec{k}-\vec{q}_R})^2} \psi_{\vec{k}} (R)
\end{equation} 
where $\hat{T}_{\vec{r}_0}$ is the translation operator in real space 
$\hat{T}_{\vec{r}_0}  \psi (R) = \psi({\vec{r}_1} - \vec{r}_0, ..., {\vec{r}_N} - \vec{r}_0)$
, and $\vec{r}_0 = r_0(\cos\theta_r, \sin\theta_r,0)$, $\vec{q}_R = q_R(\cos\theta_q, \sin\theta_q,0)$.
 By taking $r_0$ of the order of 2 \AA~, such as the location of the first maximum of $\rho (r )$, $\pi (q)$ shows two main peaks, one centered around $q = 0$, and one at $q_R$. This suggests the more systematic and Fourier--space based approach of considering  multimodal packets (the new scheme also allows a more efficient exploration of parameters):
 \begin{equation}
\pi_{ \{A_i,\sigma_i,q_i \}} (q) = \sum_{i=1}^{n} A_i f_i(\frac{q-q_i}{\sigma_i})
\end{equation}
where A can be positive or negative, $n$ is the number of peaks, and $f_i$ is chosen to be a Gaussian or a Lorentzian. Since the normalization of the wavefunction has already to be taken into account by the Monte Carlo computation, $\pi$ is defined but for a multiplicative constant, that is we fix $A_2 = 1$.
Taking $\beta$  independent of $q$,  the density profile function for a given packet can be computed by means of Monte Carlo, where in eq. (\ref{eq:psi_shadow}) one replaces 
$\frac{1}{\sqrt{N_q}} e^{i\vec{q}\cdot \tilde{\vec{s}}_j} 
\to  \eta(|\tilde{\vec{s}}_j|) = \int d^2 q \frac{\pi(q)}{\sqrt{N_q}} e^{i\vec{q}\cdot \tilde{\vec{s}}_j} $. Since results for $\beta=0$ and $\beta=0.3$ differ only slightly (and keeping in mind the qualitative nature of our goals), we can consistently take $\beta$ fixed.

Since taking $f_i$ Lorentzian or Gaussian, and adding a third peak does not make a significant difference, we reported in the main text plots for packets in the form:
\begin{equation}
\pi_{ \{A_1,\sigma_1,q_2, \sigma_2 \} } (q) = A_1 e^{-\frac{q^2}{2 \sigma_1^2}} + e^{-\frac{(q-q_2)^2}{2 \sigma_2^2}} \label{bigauss}
\end{equation}
see {Fig.~\ref{fig:profileS} here for more quantitative -- direct and momentum space -- information.}

\section{D) Discussion }

Some profiles are shown in Fig.~\ref{fig:profileS} and outline the following scenario: phonons are unable to excite significant density modulations (orange solid line), as can be understood also from the static density response function; maxons can create density modulations with typical wavelengths of $q \sim 1.5$\AA$^{-1}$
(red dashed); rotons in combination with low-energy phonons create density modulations with amplitudes and wavelengths similar to the vortex, at least in the $q \sim 2.$\AA${}^{-1}$ region (blue and cyan dashed--dotted); adding a third high-$q$ phonon peak screw these packets up (green dotted).

Notice that if the Fourier trasforms reported are in good qualitative agreement with the vortex one, none of the built profiles has a significant contribution at $q< 1$\AA, unlike to the vortex case.
{The likely reason for this is that we are expanding in the single-excitation subspace, 
and not in the complete Fock basis $| \{ n_{\vec{q}} \}_{\vec{q}} \rangle$. 
Since the modulus of the many-body function (leaving out the phase) is a bosonic many-body wavefunction, 
it is in principle possible to reproduce any feature of the density profile with a full Fock expansion. 
It seems that several phonons and/or a few multirotons with small (total) momentum are needed in order to retrieve the short-wavelength structure. 
Finally we remark that, in our single-excitation approach, the amplitude of density modulations 
depends on the inverse of $L_z$, the box size along the vortex axis, i.e. on the density of excitations per unit length; this is shown in Fig. \ref{fig:N_Scaling}.}
This has allowed us to estimate the number of virtual rotons needed to reproduce the vortex density 
profile as discussed in the main text.

In spite of the limitation of this single--excitation approach,
we are more interested here in the short distance behavior (large $q$) of the vortex 
density profile because its long distance behavior is not much relevant for the reconnection event. 
In fact, as two vortices approach each other both the phase and the modulus of the wave function 
will change with respect to those of two independent vortices. This intervortex effect will start 
from large distances so that by the time the two vortices reconnect the large distance profile, 
that is well captured by GPE, is expected to be rather different from that of a single vortex.  

We remark that, by a suitable choice of the wave--packet's parameters one can obtain not only
depletion of the density along the vortex axis but also augmentation, with oscillation of the density
that decay at larger $r$.
This flipped profile should not surprise us: if you suppose that the time dynamics of the roton packet is similar to density oscillations, then depletions oscillate to augmentations and viceversa.
In fact, we have some preliminary numerical evidence of this fact, since in our approach one is tempted to compute the time evolution of (\ref{eq:packet}) as:
\begin{equation}
\Psi (R, t) = \int d^2 q \ e^{-i\frac{\epsilon(q)}{\hbar}t} \ \pi (\vec{q}) \psi_{\vec{q}} (R) .
\end{equation}
Unfortunately, $\psi_{\vec{q}} (R)$ is not the exact eigenvector of the Hamiltonian and,  even though the error in the average energy is small, quantum energy fluctuations could  affect sensibly the dynamics. Assessing the robustness of the time evolution against these effects (and solving some other technical problems such as the spreading of  packets in space) goes beyond the scope of this work; we refer to an eventual future publication for a full treatment of these issues.

\clearpage


\begin{thebibliography}{99}
\bibitem{ftnota} We take the chance here to point out that the Fourier transforms shown 
                 in \cite{prb14} have accidentally been plotted with an incorrect minus sign.
\bibitem{footnote1} Some of the vorticity in the lower chamber might well enter into the upper chamber but 
this should not be a problem as long as the density of vortices in the two chambers are well 
separated because the number of reconnections has a strong dependence on $L$.
\bibitem{footnote2} Notice that each ${}^4$He atom impinging on the bolometer deposits 
the full roton energy because the $E_b$ lost in the evaporation is regained once a thin 
film of the superfluid covers the bolometer as in Ref. \onlinecite{P}.
\end{thebibliography}
\end{document}